# The spin relaxation of nitrogen donors in 6H SiC crystals as studied by the electron spin echo method

D. Savchenko,[1,2,a)] B. Shanina,[3] E. Kalabukhova,[3] A. Pöppl,[4] J. Lančok,[1] and E. Mokhov[5,6]

[1]*Institute of Physics of the Czech Academy of Sciences, Prague, 182 21, Czech Republic*

[2]*National Technical University of Ukraine "Kyiv Polytechnic Institute", Kyiv, 03056, Ukraine*

[3]*V.E. Lashkaryov Institute of Semiconductor Physics, NASU, Kyiv, 03028, Ukraine*

[4]*Institute of Experimental Physics II, Leipzig University, Leipzig, D-04103, Germany*

[5]*A.F. Ioffe Physical Technical Institute, RAS, St. Petersburg, 194021, Russia*

[6]*Saint-Petersburg National Research University of Information Technologies, Mechanics and Optics, St. Petersburg, 19710, Russia*

We present the detailed study of the spin kinetics of the nitrogen (N) donor electrons in 6H SiC wafers grown by Lely method and by sublimation "sandwich method" (SSM) with a donor concentration of about $10^{17}$ cm$^{-3}$ at $T$ = 10-40 K. The donor electrons of the N donors substituting quasi-cubic "k1" and "k2" sites (N$_{k1,k2}$) in both types of the samples revealed the similar temperature dependence of the spin-lattice relaxation rate ($T_1^{-1}$), which was described by the direct one-phonon and two-phonon processes induced by the acoustic phonons proportional to $T$ and to $T^9$, respectively. The character of the temperature dependence of the $T_1^{-1}$ for the donor electrons of N substituting hexagonal ("h") site (N$_h$) in both types of 6H SiC samples indicates that the donor electrons relax through the fast-relaxing centers by means of the cross-relaxation process. The observed enhancement of the phase memory relaxation rate ($T_m^{-1}$) with the temperature increase for the N$_h$ donors in both types of the samples, as well as for the N$_{k1,k2}$ donors in Lely grown 6H SiC, was explained by the growth of the free electron concentration with the temperature increase and their exchange scattering at the N donor centers. The observed significant shortening of the phase memory relaxation time $T_m$ for the N$_{k1,k2}$ donors in the SSM grown sample with the temperature lowering is caused by hopping motion of the electrons between the occupied and unoccupied states of the N donors at N$_h$ and N$_{k1,k2}$ sites. The impact of the N donor pairs, triads, distant donor pairs formed in n-type 6H SiC wafers on the spin relaxation times was discussed.

_________________________

a) Author to whom correspondence should be addressed. Electronic mail: dariyasavchenko@gmail.com.

## I. INTRODUCTION

Nowadays, the relaxation processes in the spin systems are of great interest because of the spintronics development and perspectives for building the quantum computers on the basis of the shallow donors in semiconductors like the phosphorus donors in silicon (Si:P) or GaAs quantum dots. The reason is that the electrons bound to shallow donors have rather long spin-lattice relaxation ($T_1$) and phase memory ($T_m$) times to maintain fidelity of gate operations for a nuclear spin quantum computer. Particularly, the $T_m$ time, which is more relevant for the quantum information processing, is required to be no shorter than a few microseconds.[1] At the low donor concentrations (< $10^{16}$ P/cm$^3$), the $T_1$ time in Si:P varies from microseconds at 20 K to thousands of seconds at 2 K,[2,3] and is independent of the phosphorus concentration. The phase memory time in the isotopically purified $^{28}$Si:P was estimated to be $T_m \sim 60$ ms at 7 K.[1] At the same time, the experiments on the isolated donors in n-type silicon carbide (SiC) have shown that the shallow nitrogen (N) donors in 6H polytype of SiC have a sufficiently long $T_m$ time at temperatures higher than in Si:P.[4] It was found that the $T_m$ time exceeds 100 µs at 50 K (125; 333; 588 µs at 50 K; 40 K and 20 K respectively) for the shallow N donors substituting quasi-cubic ("k1", "k2") sites (N$_{k1,k2}$) in 6H SiC with a donor concentration ($N_D - N_A$) ≈ $1 \times 10^{17}$ cm$^{-3}$ ($N_D$ and $N_A$ are donor and acceptor concentrations, respectively),[4] according to the pulsed electron paramagnetic resonance (EPR) studies. It was expected that the $T_m$ time should be increased if the donor concentration is lowered.

While the values of the spin relaxation times were found in 6H SiC for N$_{k1,k2}$ donors, there are no data about the $T_1$ and $T_m$ times for the N donors substituting hexagonal "h" position (N$_h$) and the mechanism of relaxation times in SiC remains still poorly understood in the wide temperature range. The data available in the literature concerns mostly the concentration dependence of the spin-lattice relaxation (SLR) rate ($T_1^{-1}$) for the N donors without accounting the difference in the behavior of the SLR for $N_h$ and $N_{k1,k2}$ centers in 6H SiC.[5]

It is known that the EPR spectra of N donors in SiC are inhomogeneously broadened due to the superhyperfine interaction with $^{29}$Si and $^{13}$C nuclei[6,7] and homogenous line broadening contributions due to spin relaxation processes are difficult to extract from the continuous wave (CW) EPR line widths. In contrast to CW EPR measurements the electron spin echo (ESE) experiments provide the opportunity to measure independently both spin relaxation times $T_1$ and $T_m$. Therefore the measurements of relaxation times using ESE spectroscopy allows to identify the value of the relaxation time $T_m$ which does not depend on magnetic field inhomogeneity nor on hyperfine coupling and hence has an advantage over the EPR linewidth method.[8] Moreover, at $T < 20$ K the N donors in SiC have long $T_1$ times and the two-pulse sequence that gives an ESE is repeated faster than about once every five times the $T_1$ relaxation time, so that the spin system will not return to equilibrium between pulse sequences. In this case, the z-magnetization is decreased, less magnetization is available to project into the xy plane, and the ESE amplitude is decreased.[8] As a result, the $T_1$ time can be determined from the dependence of ESE amplitude on pulse repetition rate and if the spectral diffusion makes a constant contribution over the range of repetition times used, the measured $T_1$ approximates the actual $T_1$.[8] Thus, at $T < 20$ K the described technique has an advantage over the inversion-recovery pulsed EPR methods that are affected by spectral diffusion effects take place and saturation recovery experiments which are less subjected to spectral diffusion but are experimentally more demanding in general.[9]

In the present work we report a new comprehensive study of the temperature behavior of the $T_1$ and $T_m$ times for $N_{k1,k2}$ and $N_h$ donors in 6H SiC crystals grown by different methods in the temperature range 4.2-40 K employing the ESE phenomenon to get the detailed information about the mechanism of the relaxation processes.

## II. MATERIALS AND METHODS

The n-type 6H SiC wafers with $(N_D - N_A) \approx (1\text{-}5) \times 10^{17}$ cm$^{-3}$ were grown by Lely method and sublimation "sandwich method" (SSM).[10,11] The growth of the n-type 6H SiC wafers by modified Lely method was carried out around 2200-2400°C of the growth temperature and 30-50 mbar of Ar pressure with the growth rate of 1.2 mm/h on [0001] Si face using polycrystalline SiC as source materials. The growth of the n-type 6H SiC wafers by SSM was carried out at 1900$^0$C with the growth rate of 0.2 mm/h on the [0001]C face in a tantalum container under Si excess partial pressure using SiC micropowder with Si/C ~ 1.05 as a source of vapor composition. The size of the samples was about $7 \times 4 \times 0.3$ mm.

The CW and pulsed EPR measurements were performed on X-band (9.4-9.7 GHz) Bruker ELEXYS E580 spectrometer in the temperature range from 130 K to 10 K. The CW EPR experiments were carried out using the ER 4122 SHQE SuperX High-Q cavity, while for the pulsed EPR measurements the EN 4118X-MD4 cavity was used. The FS ESE spectra were measured using two-pulse Hahn echo sequence: $\pi/2 - \tau - \pi - \tau -$ echo with the pulse lengths: $\pi/2 = 96$ ns, $\tau = 600$ ns, $\pi = 192$ ns. The $T_1$ time of paramagnetic centers (PC) was estimated from ESE signal intensity changing under the variation of the shot repetition time,[9] while the $T_m$ time was determined from the two-pulse ESE decay.

## III. EXPERIMENTAL RESULTS AND ANALYSIS

### A. The temperature behavior of the CW EPR and FS ESE spectra in 6H SiC crystals

Fig. 1 shows the X-band CW EPR spectra of N donors observed in Lely grown 6H SiC and in 6H SiC grown by SSM in the temperature range from 60 K to 130 K. At 60 K, the EPR spectrum consists of two overlapping triplet lines from $N_{k1}$ ($g_{\parallel}$ = 2.0040(3), $g_{\perp}$ = 2.0026(3), $A_{\parallel} = A_{\perp}$ = 1.20 mT) and $N_{k2}$ ($g_{\parallel}$ = 2.0037(3), $g_{\perp}$ = 2.0030(3), $A_{\parallel} = A_{\perp}$ = 1.19 mT) donors.[6] The central line of $N_{k1,k2}$ triplets coincides with the EPR spectrum of $N_h$ ($g_{\parallel}$ = 2.0048(3), $g_{\perp}$ = 2.0028(3)) representing the line with small unresolved hyperfine (hf) splitting ($A_{\parallel}$ = 0.1 mT, $A_{\perp}$ = 0.08 mT).[7] Along with the EPR spectrum from the isolated N centers, the lines of comparatively low intensity are observed in-between of $N_{k1,k2}$ triplet lines. In accordance with Ref. [6], the $N_x$ lines ($g_{\parallel}$ = 2.0043(3), $g_{\perp}$ = 2.0029(3), $A_{\parallel} = A_{\perp}$ = 0.6 mT) were attributed to the triplet center with $S = 1$ responsible for the distant donor pairs between the N atoms residing at quasi-cubic and hexagonal sites ($N_{k1,k2}Si_kN_h$). It is clearly seen from Fig. 1 that the contribution of the distant donor pairs in the EPR spectrum of the N donors in SSM grown 6H SiC is small. The detailed analysis of the temperature behavior of CW EPR spectra in both samples was undertaken in Ref. [12]. Particularly from the simulation of the ESR spectra measured in the 6H SiC samples at 60 K the relative intensity ratio $I(N_{k2}) : I(N_{k1}) : I(N_h) : I(N_x)$ was found to be: 1.0 : 1.3 : 1.9 : 0.2 for the Lely grown and 1.0 : 0.9 : 3.0 : 0.1 for the SSM grown 6H SiC samples. Thus, the intensity of the EPR line from $N_h$ donors with respect to that of $N_{k1,k2}$ triplet lines is higher in SSM grown 6H SiC than in Lely grown 6H SiC sample indicating that $N_h$ donors are mostly in isolated state in 6H SiC grown by SSM. And the intensity of $N_x$ triplet is lower in SSM grown sample than in Lely grown 6H SiC.

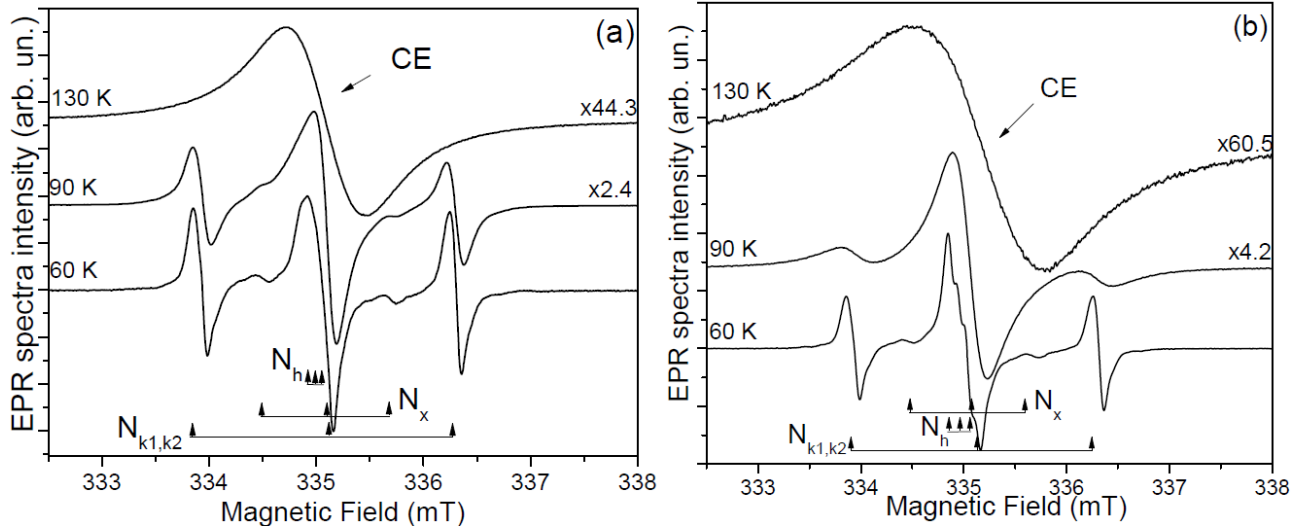

FIG. 1. The temperature behavior of the X-band EPR spectra measured in Lely (a) and SSM grown (b) 6H SiC samples measured from 60 K to 130 K, $\boldsymbol{B}\|c$. On the right side of each spectrum the multiplication unit of the EPR spectra intensity is pointed with respect to the intensity of the EPR spectra measured at 60 K that was taken equal to 1.

With the temperature increase up to 70 K, the single line of Lorentzian shape with $g_\| = 2.0043(3)$, $g_\perp = 2.0030(3)$ appears in the EPR spectrum. The emergence of the single EPR line is accompanied by the disappearance of the $N_h$ EPR line at 80 K and $N_{k1,k2}$ triplet lines at 100-110 K. In accordance with Ref. [12], the observed single Lorentzian line with the temperature-dependent linewidth and intensity was attributed to the conduction electrons (CE).

A further decrease of the temperature to 40 K gives rise to the saturation of EPR spectra in both types of the samples due to the increase of the spin relaxation times of the N paramagnetic centers and, as a result, the measurements of the $T_1$ and $T_m$ times of N centers become possible using ESE phenomenon at $T < 40$ K. It should be noted that saturation effect can be reduced by the registration of the EPR spectrum at high frequency due to the shortening of the spin-lattice relaxation time $T_1$ at high magnetic fields.[13]

Fig. 2 shows the two-pulse field sweep detected electron spin echo (FS ESE) spectra of N donors in the 6H SiC samples grown by Lely method and by SSM in the 40-10 K range.

Comparing the temperature behavior of the intensity ratio between the $N_h$ and $N_{k1,k2}$ FS ESE signals in both samples, one can see that it has a different character. The intensity ratio between $N_h$ and $N_{k1,k2}$ FS ESE signals remains unchanged in the Lely grown 6H SiC sample in the whole temperature interval, while in 6H SiC grown by SSM it varies significantly between 40 and 10 K. This indicates that in contrary to the Lely grown 6H SiC sample, the spin relaxation times for $N_h$ and $N_{k1,k2}$ donors in 6H SiC grown by SSM have different temperature behavior.

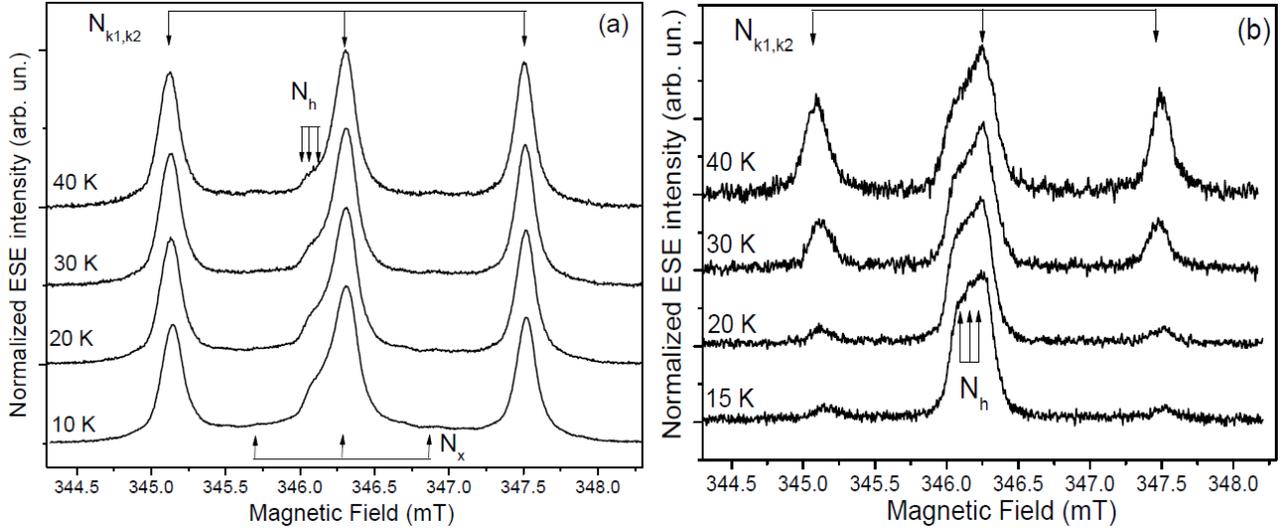

FIG. 2. The temperature behavior of the FS ESE spectra measured in Lely grown (a) and grown by SSM (b) 6H SiC. ***B***||***c***. The shot repetition time for Lely grown sample was set to 12.24 ms, for SSM grown sample it was 10.2 ms at $T = 40$-$20$ K and 40.8 ms at $T = 15$ K.

**B. The temperature dependence of spin relaxation times of N donor electrons in 6H SiC sample grown by Lely method and by SSM**

*1. The temperature dependence of spin-lattice relaxation time*

Fig. 3 shows the temperature dependence of the SLR rate in the temperature interval from 10 K to 40 K for donor electrons of $N_{k1,k2}$ and $N_h$ in the 6H SiC samples grown by Lely method and by SSM. As was shown in Fig. 3a, the SLR rate for $N_{k1,k2}$ donors increases continuously with the temperature increase in both 6H SiC samples and can be described by the classical spin-phonon interaction (under the condition $T \ll \theta_D$, where $\theta_D$ is the Debye temperature for crystal) as a relaxation process via the acoustic phonons:

$$T_1^{-1}(T) = w_1 T + w_2 (T/\Theta_D)^9, \qquad (2)$$

where $T$ is the temperature (in K), $w_{1,2}$ are the relaxation rates in the one-phonon direct process and in two-phonon process for spins in Kramers doublet states, respectively.

The solid lines in Fig. 3a show the calculated $T_1^{-1}$ values using Eq. (2) that are well fitted with the experimental data using the parameters given in Table I. Analyzing the data in Table I, one can see that the SLR rate for $N_{k1,k2}$ donors have close values in both types of the samples.

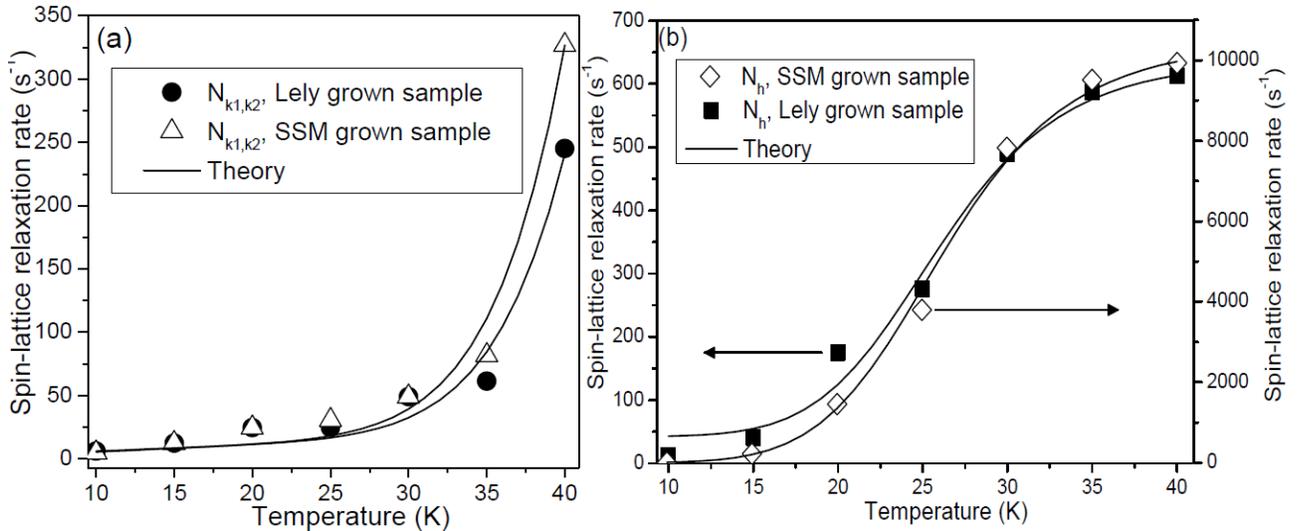

FIG. 3. The temperature dependence of SLR rate for the $N_{k1,k2}$ (a) and $N_h$ (b) donors in 6H SiC samples grown by Lely method and by SSM measured in the 40-10 K temperature range. Dots are experimental data, solid lines are the fitting using Eq. (2) (a) and Eq. (4) (b).

TABLE I. Relaxation rates of the one-phonon ($w_1$) and two-phonon ($w_2$) processes in 6H SiC samples grown by Lely method and by SSM obtained from the fitting of Eq. (2) with experimental data for $N_{k1,k2}$ donors shown on Fig. 3.

| Samples | $w_1$ (s$^{-1}$) | $w_2$ (s$^{-1}$) | $\theta_D$ (K) |
|---|---|---|---|
| Lely grown 6H SiC | 0.5 | $8.9 \times 10^{15}$ | 1300 |
| 6H SiC grown by SSM | 0.55 | $6 \times 10^{15}$ | 1200 |

*$\theta_D$ for 6H SiC is 1200 K.[14]

In contrast to the temperature behavior of the SLR rate for $N_{k1,k2}$ donors, the temperature dependence of the SLR rate for $N_h$ centers at higher temperatures has smooth character and becomes flat at $T \geq 40$ K (see Fig. 3b). Such temperature behavior of the SLR rate occurs when the PC relaxes via the fast relaxing (FR) spin system through the cross-relaxation process. For the $N_h$ donors, which have shallow energy levels and a larger radius of the wave function in comparison with $N_{k1,k2}$ donors, the CE may play the role of the FR spin system. In this case, in accordance with Ref. [15], the SLR involves two following steps: in the first step the donor spin system transfers their energy to the FR spin system through the spin-spin exchange interaction and then, in the second step, the FR spin system relaxes fast via spin-lattice interaction. The relaxation rate is controlled by the slower step of the relaxation process and the whole process can be described by the following function:[15]

$$T_1^{-1}(T) = \frac{n_{FR} U W_{FR}}{(1 + N_d/n_{FR}) n_{FR} U + W_{FR}}, \qquad (3)$$

where $n_{FR}$, $N_d$ – are the concentrations of the FR spin system and shallow donors, respectively ($N_d/n_{FR} \gg 1$), $U$ represents cross-relaxation rate between two spin systems and does not depend on the temperature, $W_{FR}(T) = W_{FR,0} \times T^n$ is the rate of the SLR of the FR spin system, where $W_{FR,0}$ is the rate of SLR of FR system at $T = 1$ K.[15]

At low temperatures the free electrons relax via a spin-orbit mechanism and the SLR is described by the function: $W_{FR,0} \times (T)^5$. At high temperatures (when $W_{FR} \gg N_d U$) the $T_1^{-1}$ becomes equal to $n_{FR} U$ and since the CE concentration is a temperature dependent value $n(T) \sim \exp(-E/kT)$ the Eq. (3) becomes a continuously increasing function with the temperature. For this reason, the free electron spin system cannot be considered as a FR system for the $N_h$ donor.

The second candidate that can be considered as the FR centers are exchange pairs or triads formed between PC.[16] The distant donor pairs $N_x$ cannot be considered as FR centers because their spin relaxation times are comparable with that for the N donors. The experimental curves plotted on Fig. 3b for the $N_h$ donors are well described by Eq. (3) with the following function:

$$T_1^{-1}(T) = c_0 T + \frac{c_1 T^7}{c_2 + T^7}, \tag{4}$$

where $c_0$ is the intrinsic SLR of the donor electrons at $T = 1$ K; $c_1 = n_{FR} \times U$ is the cross-relaxation rate; $c_2 = N_d \times U / W_{FR,0}$; $W_{FR,0}$ can be calculated as $W_{FR,0} = (N_d/n_{FR}) \times (c_1/c_2)$.

The fitting parameters of Eq. (4) with experimental data are given in Table II. The $W_{FR,0}$ values were obtained with $n_{FR} \approx 2 \times 10^{11}$ cm$^{-3}$. The $n_{FR}$ value was estimated as the concentration of the exchange pairs of the two N donor centers, when $N_d = 10^{17}$ cm$^{-3}$. Thus, the $N_h$ donor electrons relax through the FR exchange pairs or clusters of the donor centers by means of the cross-relaxation process.

TABLE II. The fitting parameters of Eq. (4) with experimental data for $N_h$ donors.

| Samples | $c_0$ (s$^{-1}$) | $c_1$ (s$^{-1}\times$K$^{-7}$) | $c_2$ (K$^7$), $10^9$ | $W_{FR,0}$ (s$^{-1}\cdot$K$^{-7}$) |
|---|---|---|---|---|
| Lely grown 6H SiC | 12 | $6\times10^2$ | 8 | $4\times10^{-2}$ |
| 6H SiC grown by SSM | < 2 | $1.05\times10^4$ | 10.7 | $6\times10^{-1}$ |

*2. The temperature dependence of spin-spin relaxation time*

It was found that the time decay of the ESE signal amplitude for the N donor centers in 6H SiC is described by a superposition of two exponential functions:

$$I(t) = A_1 \exp(-t/T_{m,f}) + A_2 \exp(-t/T_{m,s}), \tag{5}$$

where $t$ – time, $A_1$, $A_2$ – constant values, $T_{m,f}$ and $T_{m,s}$ are fast and slow components of ESE decay, respectively.

The main contribution to the ESE decay comes from the slow exponent $T_{m,s}$, while the exponent with on the order of magnitude faster time $T_{m,f}$ gives the negligible small contribution to the total exponential decay of the ESE signal for the $N_h$ donors in both types of the samples, as well as for the $N_{k1,k2}$ donors in Lely grown 6H SiC. In contrast, the main contribution in the time decay of the ESE signal for the $N_{k1,k2}$ donors in 6H SiC grown by SSM is single exponential with a short decay constant $T_{m,f}$. As it is seen from Fig. 4, the temperature dependence of $T_{m,s}^{-1}$ and $T_{m,f}^{-1}$ rates extracted from the ESE time decay curves using Eq. (5) for the $N_{k1,k2}$ donors 6H SiC grown by Lely method has an opposite character: the $T_{m,s}^{-1}$ grows while $T_{m,f}^{-1}$ decreases with the temperature increase.

Recording the curves of the ESE time decay at different temperature points, one can obtain the temperature dependence of the phase memory relaxation rate ($T_m^{-1}$) for the N donors in both samples shown in Fig. 5. As can be seen from Fig. 5a, the enhancement of the $T_m^{-1}$ with the temperature increase was observed for the shallow $N_h$ donors in both samples as well as for the $N_{k1,k2}$ donors in Lely grown 6H SiC crystal. At the same time, in 6H SiC grown by SSM, the $T_m^{-1}$ for the $N_{k1,k2}$ exhibits a dramatic increase with the temperature lowering (see Fig. 5b).

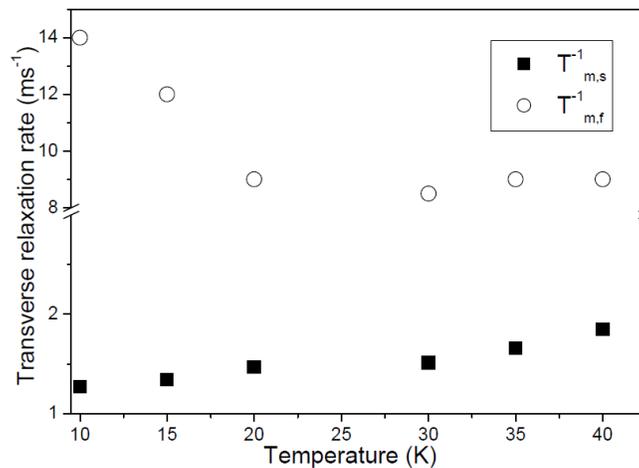

FIG. 4. The temperature dependence of $T_{m,s}^{-1}$ and $T_{m,f}^{-1}$ rates extracted from the ESE time decay curves for $N_{k1,k2}$ in Lely grown 6H SiC sample using Eq. (5).

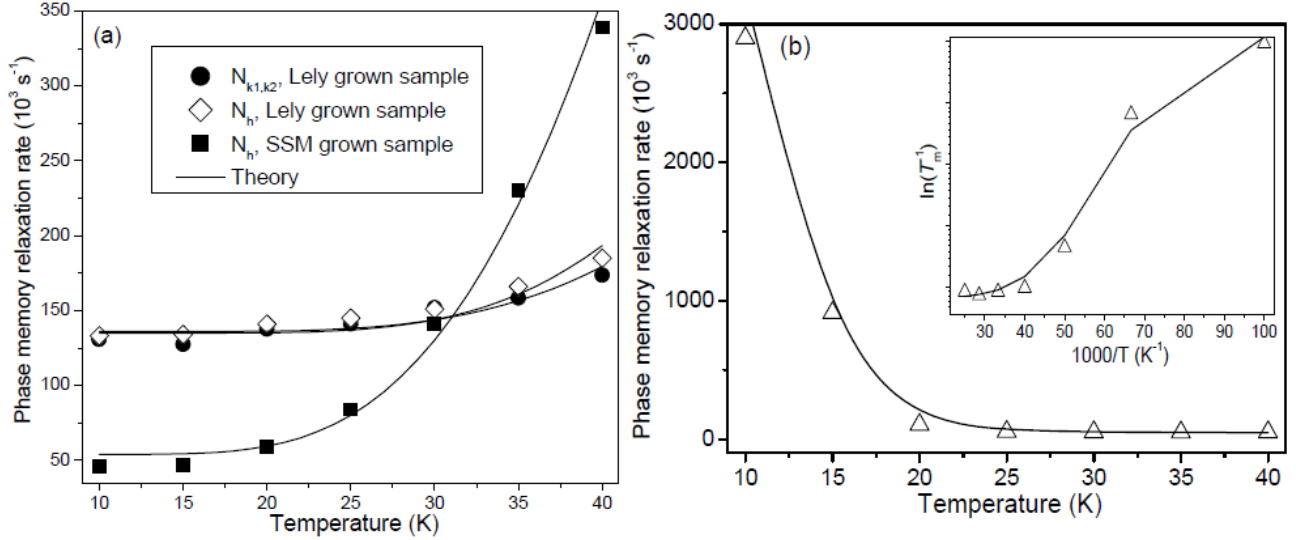

FIG. 5. The temperature dependence of $T_m^{-1}$ for the $N_{k1,k2}$ and $N_h$ donors in Lely grown 6H SiC, for the $N_h$ donors in 6H SiC grown by SSM (a) and for the $N_{k1,k2}$ donors (b) in SSM grown 6H SiC. The inset on Fig. 5(b) shows the temperature dependence of $T_m^{-1}$ in the logarithmic scale.

In principle, the $T_m^{-1}$ value should be temperature-independent because it is determined by the local fields of spin-spin interactions of the PC. Therefore, the observed temperature dependence of the $T_m^{-1}$ indicates that there is a significant spin-coupling between N donors and other spin system, which has a temperature-dependent behavior. Among such spin systems, the free electrons, having a temperature-dependent concentration value $n$, are the most suitable candidates. In this case, the $T_m^{-1}$ is determined by the exchange interaction between N donors and free electrons, governing by the spin flip-flop rate, and will depend on the concentration of the free electrons $n$.

In the case of the weak compensation ($N_D \gg N_A$) the free electron concentration $n$ as a function of the temperature is known as:

$$n(T) = \left((N_D - N_A)N_c\right)^{1/2} \exp(-\Delta E_D/2kT), \tag{6}$$

where $N_c = 2.8 \times 10^{15} \times T^{3/2}$ is a number of electron states in the conduction band for 6H SiC; $\Delta E_D = E_c - E_D$ is the distance between the conduction band bottom and the donor energy level. The free electron concentration in the conduction band ($n$) estimated from Eq. (6) varies from $1.4 \cdot 10^{11}$ cm$^{-3}$ to $2 \cdot 10^{13}$ cm$^{-3}$ with the temperature increase from 10 K to 40 K.

The spin-spin relaxation rate $T_m^{-1}$ due to exchange scattering can be written as following:[15]

$$T_m^{-1}(T) = T_{m,0}^{-1} + U \times n(T). \tag{7a}$$

Substituting Eq. (6) for $n(T)$ into Eq. (7a), one can get an expression for the temperature dependence of $T_m^{-1}$, which describes well the experimental data for the $N_h$ and $N_{k1,k2}$ donors in Lely grown 6H SiC and for the $N_h$ donors in 6H SiC grown by SSM samples shown in Fig. 5a:

$$T_m^{-1}(T) = T_{m,0}^{-1} + \frac{w_2}{1 + (1 + 8 \cdot (N_D / N_{c,0}) \cdot (300)^{3/2} \cdot \exp(\Delta E_D / kT) / T^{3/2})^{1/2}}, \tag{7}$$

where $T_{m,0}^{-1}$ is the intrinsic phase memory relaxation rate of the (isolated) N donor spin; $N_{c,0} = 2.5 \times 10^{19} \times (m^*/m)^{3/2}$, $w_2 = 2(N_D - N_A) \times U$ is the spin-spin relaxation rate caused by exchange interaction between donor and free electron spins. The results of fitting of Eq. (7) with experimental $T_m^{-1}$ values for the $N_h$ and $N_{k1,k2}$ donors in Lely grown 6H SiC and for the $N_h$ donors in 6H SiC grown by SSM are shown on Fig. 5a. The corresponding fitting parameters are represented in Table III.

TABLE III. $T_{m,0}^{-1}$, $w_2$ and $\Delta E_D$ parameters for $N_h$ and $N_{k1,k2}$ donors in Lely grown 6H SiC and $N_h$ donors in 6H SiC grown by SSM obtained from the fitting of the experimental data with Eq. (7).

| Samples | Donor | $T_{m,0}^{-1}$ ($10^5$ s$^{-1}$) | $w_2$ ($10^8$ s$^{-1}$) | $\Delta E_D$ (meV) | $N_D$, ($10^{17}$ cm$^{-3}$) |
|---|---|---|---|---|---|
| Lely grown 6H SiC | $N_h$ | 1.40 | 1.5 | 55.7 | 1.32 |
|  | $N_{k1,k2}$ | 1.35 | 1.4 | 55.7 | 1.32 |
| 6H SiC grown by SSM | $N_h$ | 0.5 | 0.45 | 32 | 2.4 |

Thus, the obtained data show that the exchange scattering plays an important role in the spin-spin relaxation process for the N donor spins. The $\Delta E_D$ values in Table III represent the distance between the bottom of the conduction band and the energy level of the shallowest donor $N_h$, which mostly contributes to the free electrons into the conduction band. The reduction of the activation energy values compared to that accepted for the $N_h$ donors (80 meV)[17] can be explained by the formation of an impurity band when the concentration of N donors becomes high enough. In this case, the impurity band approaches the conduction band and the ionization energy of N decreases because the upper edge of the impurity band for N donors begins to play the main role in the thermal ionization process of the donor electrons.

As was shown in Fig. 5b, in 6H SiC grown by SSM the $T_m^{-1}$ for the $N_{k1,k2}$ donors (in contrast to $N_h$) exhibits a dramatic enhancement with the temperature decrease and the main contribution to the time decay of the ESE signal amplitude for $N_{k1,k2}$ comes from the exponent with the $T_{m,f}$ time. The increase of the $T_m^{-1}$ for $N_{k1,k2}$ donors can be explained by the influence of the hopping conduction process that occurred in 6H SiC grown by SSM. A small contribution of the $T_{m,f}^{-1}$ in $T_m^{-1}$ for the $N_{k1,k2}$ and $N_h$ donors in Lely grown 6H SiC was likewise explained by the small effect of the hopping conduction process in this sample.

It should be noted that the known phenomenon of the "motional" narrowing of EPR linewidth (the latter is a reciprocal spin-spin relaxation time)[18] is caused by the hopping motion of electrons with the variable length of jumps[19] within the system of N donors and exchange interaction between the N donors and CE.[12,20]

For the $N_{k1,k2}$ donors in 6H SiC grown by SSM the electron hopping motion induces the fluctuation of the local magnetic field at the donor site due to the exchange coupling between N donors and free electrons. Following Mott,[19] the fluctuation frequency is equal to the probability of the electron hopping between different sites:

$$\tau_h^{-1}(T) = v_{ph} \exp\left((-T_0/T)^\lambda\right) \qquad (8)$$

with $\lambda = 1/4$, 1/3, or 1/2; $v_{ph}$ – maximum of the acoustic phonon frequency (18.9 THz in 6H SiC).[19]

On the other hand, the randomly fluctuating magnetic field causes the spin diffusion due to the spin-spin interaction between N donors and free electrons, which leads to the random excitation of the spin system. In this case, according to the theory of the time evolution of spin density matrix, the kinetic equations for the transversal magnetization contain $T_m^{-1}$ (the inverse time for relaxation of transversal component of magnetization), which is expressed through the Fourier transforms $j(\omega - \omega_0)$, $j(\omega)$ of the correlation functions $c_h$ of the local magnetic field fluctuations:[18]

$$M_\pm \times i(\omega - \omega_0) = -i\omega_1(M_z - M_0) + (j(\omega - \omega_0) + j(\omega)) \times M_\pm, \qquad (9)$$

where the real part of the right side of Eq. (9) plays the role of the inverse relaxation time $T_m^{-1} = j(\omega - \omega_0) + j(\omega)$, $\omega$ and $\omega_0$ are the MW field and resonance magnetic field frequencies, respectively, $\omega_1$ is the amplitude of the MW field, $M_z$ is the equilibrium value of z-component of the magnetization.

The correlation function of the local field fluctuations can be represented by the exponentially decaying function with the correlation time $\tau_h \ll T_m$:[18]

$$c_h(t) = \gamma^2 <h_L \times (t+\tau)h_L \times (t)> = <\gamma^2 h_L^2> \times \exp(-t/\tau_h) \qquad (10)$$

where $\gamma = (g\mu_B/\hbar)$; $<h_L^2>$ – an average square of the exchange local field, when the exchange interaction between free electron spin system and donor spins is written as $\hat{H}_{ex} = \mathbf{h} \cdot \mathbf{M}$.

From Eq. (10) it follows that:

$$j(\omega - \omega_0) = \frac{<\gamma^2 h_L^2> \tau_h}{1+(\omega-\omega_0)^2 \tau_h^2} \quad (11)$$

where $\tau_h$ is a correlation time of the electron hopping motion.

As a result, the experimental curve shown in Fig. 5b can be described by the expression:

$$T_m^{-1}(T) = T_{m,0}^{-1} + \frac{w_h \exp\left((T_0/T)^\lambda\right)}{1+(\omega-\omega_0)^2 w_h^2 \exp\left(2(T_0/T)^\lambda\right)} \quad (12)$$

with following parameters $T_{m,0}^{-1} = 4.5 \times 10^4$ s$^{-1}$; $w_h = 0.1$ s$^{-1}$; $T_0 = 3700$ K; $\lambda = 1/2$. Comparing the obtained values with those given in Table III, it is seen that the intrinsic lifetime of the N$_{k1,k2}$ donor electron spins in the 6H SiC grown by SSM is longer (22 μs), than that (7.4 μs) in Lely grown 6H SiC.

## IV. CONCLUSIONS

The mechanisms of spin-lattice and spin-spin relaxation times for the N donor electrons substituting quasi-cubic ("k1","k2") and hexagonal ("h") positions in n-type 6H SiC wafers grown by Lely method and SSM with $(N_D - N_A) \approx 10^{17}$ cm$^{-3}$ have been studied by pulsed EPR spectroscopy in the 10-40 K temperature range.

For N$_{k1,k2}$ donors in both types of the samples, the temperature dependence of SLR in the 10-40 K temperature range was described by the direct one-phonon and two-phonon relaxation processes via acoustic phonons proportional to $T$ and to $T^9$, respectively, and it exceeds 204 ms in 6H SiC grown by SSM and 163 ms and Lely grown 6H SiC at 10 K. The temperature behavior of the $T_1^{-1}$ for donor electrons of $N_h$ in both samples was explained by the cross-relaxation process between $N_h$ and fast-relaxing centers. The $T_1$ time for $N_h$ donors at 10 K becomes by three times longer in 6H SiC grown by SSM ($T_1$ = 244 ms) as compared with that found in Lely grown 6H SiC ($T_1$ = 82 ms).

The observed enhancement of the $T_m^{-1}$ with the temperature increase for the $N_h$ in both samples and $N_{k1,k2}$ in Lely grown 6H SiC samples was explained as being caused by the exchange scattering of free electrons, which has a temperature-dependent concentration, on the N donors.

In contrast to the $N_h$ donors, the time decay of the ESE signal amplitude for the $N_{k1,k2}$ donors in 6H SiC grown by SSM has a single-exponential behavior. The key new result is that the decay rate of the $T_m^{-1}$ strongly increases with the temperature lowering. Such behavior was explained by the hopping motion of electrons between occupied and unoccupied sites of the $N_{k1,k2}$ and $N_h$ centers with the variable jump length occurred in 6H SiC grown by SSM. At the same time with the increase of the temperature up to 40 K, when the low temperature hopping conduction process is not significant anymore, the spin decoherence time for the $N_{k1,k2}$ donors becomes comparable ($T_m$ = 20 μs) with that for the $N_h$ donors at 10 K ($T_m$ = 21.79 μs) in 6H SiC grown by SSM. Thus, the presence of the hopping conduction process leads to the different spin dynamic process for the $N_h$ and $N_{k1,k2}$ donors in 6H SiC. As a result a small contribution of the $T_{m,f}^{-1}$ in $T_m^{-1}$ for $N_{k1,k2}$ and $N_h$ donors in Lely grown 6H SiC was explained by only a minor hopping conduction process in this sample. The drastic reduced hopping conduction in Lely grown 6H SiC can be explained by the formation of the distant donor pairs between $N_h$ and $N_{k1,k2}$ donors, which reduce the concentration of the isolated N donors and prevent the hopping motion of the donor electrons between occupied and unoccupied sites of the $N_{k1,k2}$ and $N_h$ centers. On the other side the presence of the distant donor pairs between $N_h$ and $N_{k1,k2}$ donors gives rise to the decrease of the spin lattice relaxation and spin decoherence time in Lely grown 6H SiC. Thus, following the obtained results, we are able to conclude that the longer spin decoherence time for N donors may be achieved by reduction of the concentration of the donor clusters (pairs, triads, distant donor pairs) formed in SiC wafers during the growth. Therefore, from the technology point of view the n-type 6H SiC wafers grown by SSM look more perspective for realizing the potential of N donors as a mainstay donor spin system in SiC for the engineering of spin qubits.

## ACKNOWLEDGMENTS

The work was supported by GA ČR 13-06697P, MEYS SAFMAT LM2015088 and LO1409 projects.